\documentclass[conference]{IEEEtran}
\IEEEoverridecommandlockouts
\usepackage{cite}
\usepackage{amsmath,amssymb,amsfonts}
\usepackage{algorithmic}
\usepackage{graphicx}
\usepackage{textcomp}
\usepackage[table]{xcolor}
\usepackage{enumitem}
\usepackage{booktabs}
\usepackage{hyperref}
\usepackage{subcaption}
\usepackage{multirow}
\usepackage{balance}

\usepackage{xurl}

\usepackage{tikz}
\usetikzlibrary{positioning,calc,arrows.meta}
\usetikzlibrary{shapes.geometric,shapes.misc}

\newcommand{\docfold}[1]{%
\fill[gray!15]
  ([xshift=1.5mm,yshift=1.5mm]#1.south east)
  --
  ([xshift=-1.5mm,yshift=1.5mm]#1.south east)
  --
  ([xshift=-1.5mm,yshift=-1.5mm]#1.south east)
  -- cycle;
\draw[gray!90]
  ([xshift=1.5mm,yshift=1.5mm]#1.south east)
  --
  ([xshift=-1.5mm,yshift=1.5mm]#1.south east)
  --
  ([xshift=-1.5mm,yshift=-1.5mm]#1.south east)
    --
  ([xshift=1.5mm,yshift=1.5mm]#1.south east);
}

\def\BibTeX{{\rm B\kern-.05em{\sc i\kern-.025em b}\kern-.08em
    T\kern-.1667em\lower.7ex\hbox{E}\kern-.125emX}}
\begin{document}

\newif\ifrevise

\ifrevise
    \newcommand{\revise}[1]{{\color{blue} #1}}
\else
    \newcommand{\revise}[1]{#1}
\fi

\newcommand{\suspend}[1]{}

\title{LLM Agents as Resilience Engineers for\\Scientific Applications}


\author{\IEEEauthorblockN{Hai Duc Nguyen\IEEEauthorrefmark{1}, Tekin Bicer\IEEEauthorrefmark{1}, Kyle Chard\IEEEauthorrefmark{2}, Ian Foster\IEEEauthorrefmark{1,2}, Bogdan Nicolae\IEEEauthorrefmark{1}}
\IEEEauthorblockA{\IEEEauthorrefmark{1} Argonne National Laboratory,
\IEEEauthorrefmark{2} University of Chicago\\
\{hai.nguyen,tbicer,bnicolae\}@anl.gov, \{chard,foster\}@uchicago.edu}}

\maketitle

\thispagestyle{plain}
\pagestyle{plain}

\begin{abstract}
Efficient checkpoint/restart support is essential for resilient HPC scientific applications, but implementing it requires substantial expertise: developers must identify recoverable state, choose globally consistent checkpoint points, and preserve application invariants during restart. We study whether frontier LLM coding agents can automate this process. We build a benchmark suite of 16 MPI applications spanning diverse domains, code sizes, and critical-state structures, and evaluate them with a no-human-in-the-loop generate--validate--revise pipeline for checkpoint/restart synthesis. Across the benchmark, the pipeline produces 41 working resilient implementations. Our results show that agent-driven resilience engineering is practical when critical state is visible or accessible through coherent abstractions: successful runs finish in under one hour on average, consume about 15M tokens, and produce implementations with negligible failure-free overhead and recovery efficiency comparable to human-written code. However, modularized and fragmented state remains a major limitation, with some failed attempts consuming over 100M tokens and 300 minutes without producing a working implementation.
\end{abstract}


\section{Introduction}
\label{sec:intro}

Modern scientific applications often execute for hours or days on large-scale HPC systems, making them vulnerable to hardware, software, and infrastructure failures whose likelihood increases with scale and runtime~\cite{cappello2014toward,Diaspora-Overview-eScience24}. Resilience is therefore essential for preserving progress. The dominant approach is checkpoint/restart, where the application periodically saves its execution state to durable storage and, after a failure, resumes from the latest checkpoint instead of restarting from scratch. However, implementing efficient checkpoint/restart support remains challenging. Developers must identify the proper critical state needed for recovery, choose checkpoint points where the distributed state is globally consistent, and implement restart logic that preserves communication and data-structure invariants. Performing these tasks efficiently requires significant engineering effort, application knowledge, and time.

Recent advances in large language models (LLMs) and tool-using coding agents~\cite{opencode,claude}, whose performance on real-world software engineering has improved rapidly~\cite{jimenez2024swebench, martinez2025dissecting}, raise a natural question: \textit{can expert resilience-engineering tasks be automated?} Checkpoint/restart integration appears well matched to coding-agent capabilities because its high-level workflow is straightforward: first identify critical state, insert checkpoint calls, then implement restart and recovery logic. In principle, coding agents could provide a new form of transparent checkpointing (e.g., DMTCP \cite{dmtcp}) by generating application-level checkpoint/restart code that captures only the critical state needed for recovery, rather than snapshotting the entire process. Realizing this promise, however, remains difficult because scientific applications vary widely in data layout, communication patterns, build systems, runtime behavior, and checkpoint/restart requirements.

This paper investigates the operational viability of \emph{agent-driven resilience engineering} as a building block for distributed HPC applications. We construct a benchmark suite of 16 MPI applications spanning diverse domains, code sizes, computation and communication patterns, and critical-state structures. We classify these applications into four groups (e.g., plain static, plain dynamic, encapsulated, and modularized) that capture increasing difficulty in identifying and restoring the state needed for recovery. We then build a no-human-in-the-loop generate--validate--revise pipeline to test whether frontier LLM coding agents can synthesize efficient checkpoint/restart implementations for these applications. Our results show that agent-driven resilience engineering is practical for a meaningful class of applications: agents reliably generate working implementations when critical state is directly visible or accessible through coherent framework abstractions, and the generated code achieves negligible failure-free overhead while mitigating injected failures with efficiency comparable to human-engineered checkpoint/restart implementations. However, our study also exposes clear limitations. Applications whose critical state is highly modularized and fragmented across subsystems remain difficult for current agents, largely because they struggle to infer a complete and correct view of the application state needed for recovery.
The contributions of this paper are as follows.

\begin{itemize}[topsep=0pt,itemsep=0pt,leftmargin=*]

\item \textbf{Critical-state classification for resilience refactoring.}
We introduce a four-class taxonomy that captures how critical state visibility, mutability, abstraction, and fragmentation affect checkpoint/restart support (\S\ref{sec:suite}).

\item \textbf{Automatic resilience-generation pipeline.}
We design a no-human-in-the-loop generate--validate pipeline that drives LLM-backed coding agents to refactor MPI applications, producing 41 working resilient implementations across 48 application--model settings (\S\ref{sec:setup}).

\item \textbf{Comprehensive evaluation and practical insights.}
We evaluate three frontier LLM models on 16 representative MPI applications.
The results show that agent-driven resilience engineering is practical: agents produce implementations with negligible failure-free overhead and recovery efficiency comparable to human-written code in less than one hour, consuming about 15M tokens on average. We also identify modularized and fragmented critical state as a key limitation for current agents (\S\ref{sec:findings}).

\end{itemize}

\section{Background and Motivation}
\label{sec:motivation}

\medskip
\noindent\textbf{Resilient scientific applications.}
Scientific applications are growing in scale, more data-intensive, often relying on high-rate instruments, complex infrastructure, storage systems, networks, and supercomputers during execution~\cite{vescovi2022linking}.
As system size and runtime increase, the probability of failure rises sharply, making resilience a first-order concern for long-running HPC applications~\cite{cappello2014toward,ferreira2011evaluating,schroeder2006large, hukerikar2016rolex}. This paper focuses on \emph{fail-stop failures}~\cite{fail-stop}, in which failures cause a process, node, or parallel task to stop executing. Such failures are particularly disruptive for scientific applications, which are often tightly coupled through the Message Passing Interface (MPI)~\cite{mpi1994standard}, a central programming model for scientific computing~\cite{bernholdt2020survey}. Without explicit fault-tolerance support, fail-stop failures typically abort the entire job, thereby losing \textit{all} computation progress.
Checkpoint/restart is the dominant approach for mitigating fail-stop failures~\cite{egwutuoha2013survey}. The idea is simple: applications periodically save their \textit{critical state} as a \textit{checkpoint} to persistent storage (e.g., a parallel file system~\cite{bent2009plfs, veloc}) and after failure, they restart from the latest checkpoint instead of redoing from scratch, converting catastrophic loss of progress into bounded rollback.

Although conceptually simple, implementing an \emph{efficient} checkpoint/restart solution remains difficult. \textit{First}, checkpointing is tightly coupled to application semantics. Developers must identify the critical state that represents progress, distinguish it from the reconstructable state, and determine safe time windows where the distributed state is consistent to save/restore that critical state. These decisions require detailed knowledge of how data structures evolve over time and how invariants are maintained across computation, communication, and I/O.
\textit{Second}, checkpoint/restart efficiency depends on ever-changing deployment-specific conditions. For example, the Young/Daly formula~\cite{young1974first,daly2006higher} relates the optimal checkpoint period to
failure rate and checkpoint I/O overhead, both of which vary across platforms, workloads, and system load.
Production HPC systems also experience performance variability caused by factors such as network congestion and interference from other jobs ~\cite{jha2020measuring,patke2021delay}. Consequently, a checkpoint/restart implementation that is efficient on one platform, scale, or workload phase may become inefficient after the input changes or the execution environment shifts. Maintaining resilience, therefore, requires not only high-quality checkpoint/restart code, but also repeatedly adapting and validating that code as applications and infrastructures evolve.

\medskip
\noindent\textbf{LLM-driven coding agents.}
Recent LLM-backed coding tools suggest a new opportunity for reducing this engineering burden. Real-world use already shows substantial adoption: GitHub reported that Copilot generated 46\% of developers' code across programming languages for its users~\cite{github2023copilot, dohmke2023economic}. More recently, coding agents, such as OpenCode~\cite{opencode}, have moved beyond single-shot code completion by iteratively and concurrently reading source files, editing repositories, executing tests, and using tool feedback to solve algorithmic problems~\cite{li2022alphacode, shinn2023reflexion}, fix software issues~\cite{yang2024sweagent}, and refine code through automated search and fault localization~\cite{zhang2024autocoderover}.

Checkpoint/restart integration has a structure that appears well matched to these capabilities. The high-level resilience strategy is often known: identify critical state, insert checkpoint logic, implement restart logic, and refine the implementation for efficiency. The difficulty lies in the low-level engineering required to realize this strategy for each application, data structure, build system, and runtime environment. A coding agent that can inspect source code, interpret build and runtime failures, modify multiple files, and iterate through validation could therefore act as a resilience-integration assistant, shifting part of the work from manual implementation toward specification-driven synthesis and testing.

However, whether this promise extends to end-to-end resilience engineering for scientific applications is unknown. Existing Agent-for-HPC studies largely evaluate narrower tasks, such as HPC function completion, OpenMP pragma generation, performance-change modeling, or MPI code generation~\cite{nichols2024parallel,schneider2024mpirigen, nichols2024hpccoder}. These tasks are valuable, but they do not capture the full workflow required to retrofit checkpoint/restart into real scientific applications, where the agent must reason about distributed state, consistency, restart semantics, build systems, runtime behavior, and performance overhead. Moreover, passing tests is not sufficient evidence of correctness. Many automated repair systems can pass tests by exploiting weaknesses in the test process rather than correctly solving the underlying problem~\cite{qi2015patch}.

\section{Related Work}

As discussed above, the difficulty of implementing checkpoint/restart has motivated a substantial body of work on improving resilience efficiency in HPC. Mature libraries such as VeloC~\cite{veloc}, FTI~\cite{fti}, and SCR~\cite{scr} rely on the developer to identify critical state, place checkpoint calls at safe points, and implement restart logic. By saving only the state needed for recovery, these systems significantly reduce checkpoint size, runtime overhead, and storage cost. They also support \emph{multi-level resilience strategies}~\cite{scr,VELOC-MASCOTS21}, where checkpoints are first staged to fast node-local storage and later flushed to shared storage for stronger failure protection. A large body of work further optimizes this pipeline through asynchronous flushing~\cite{DataStatesLLM-HPDC24,GPUPrefetch-HPDC23}, write aggregation~\cite{VELOC-FGCS24}, differential and incremental checkpointing~\cite{AICkptHPDC13,GPUDedup-ICPP23}, compression and lossy encoding~\cite{SZ-CLUSTER19,LineageComp-HIPC23}, and checkpoint-cadence optimization using the Young~\cite{young1974first}--Daly~\cite{daly2006higher} formulas or adaptive variants~\cite{MLCkpt20}. These techniques substantially reduce resilience cost at scale, but they assume that the most difficult application-specific step has already been done: someone must determine what state to save, where it is safe to save it, and how to restore it correctly.

Transparent checkpointing takes the opposite approach by reducing developer involvement and effort. Systems such as BLCR~\cite{blcr}, DMTCP~\cite{dmtcp}, and CRIU~\cite{criu} snapshot the whole running process rather than developer-identified critical state \cite{adam2019checkpoint}. This improves usability, but it often captures caches, temporary buffers, and reconstructable state along with the true recovery state, leading to much larger checkpoints, weaker scalability, and portability challenges across heterogeneous environments \cite{fu2024autocheck,cao2014transparent,adam2019checkpoint}. Compiler-assisted checkpointing~\cite{bronevetsky2003, fu2024autocheck} narrows this gap through static analysis, but requires language- and compiler-specific support and can struggle with the dynamically allocated, templated, and irregular data structures common in modern HPC and AI applications.

Existing approaches, therefore, tend to offer either low developer effort or application-level efficiency, but not both. LLM-backed coding agents may help close this gap by shifting the application-specific instrumentation work to generate efficient checkpoint/restart code from human developers to tool-using agents. Yet to our knowledge, no prior work has systematically studied frontier LLM code generation for end-to-end checkpoint/restart automation.

\section{Agent-Driven Code Refactoring}
\label{sec:suite}

\begin{table}[t]
\centering
\caption{Benchmark suite of 16 scientific MPI applications grouped by their critical state structure and sorted by source-tree size (LoC) within each group.}
\label{tab:apps}
\footnotesize
\setlength{\tabcolsep}{3pt}
\renewcommand{\arraystretch}{1.05}
\begin{tabular}{@{}lrlp{3.6cm}@{}}
\toprule
\textbf{Applications} & \textbf{LoC} & \textbf{Lang.} & \textbf{Domain} \\
\midrule
\multicolumn{4}{@{}l@{}}{\textit{Plain Static (PS)}} \\
\midrule
PRK\_Stencil~\cite{prk_stencil}     & 1.0K  & C             & Stencil kernel              \\
HPCG~\cite{hpcg}                    & 6.2K  & C++           & Conjugate-gradient benchmark        \\
SW4lite~\cite{sw4lite}              & 57K   & C++   & Seismic wave propagation    \\
HyPar~\cite{hypar}                  & 63K   & C             & Hyper/parabolic PDE solver     \\
\midrule
\multicolumn{4}{@{}l@{}}{\textit{Plain Dynamic (PD)}} \\
\midrule
CoMD~\cite{comd}                    & 5.6K  & C             & Molecular dynamics          \\
SPPARKS~\cite{spparks}              & 52K   & C++           & Kinetic Monte Carlo         \\
CLAMR~\cite{clamr}                  & 81K   & C/C++         & Shallow-water hydrodynamics   \\
SPARTA~\cite{sparta}                & 174K  & C++           & DSMC rarefied-gas simulation        \\
LAMMPS~\cite{lammps}                & 627K  & C++           & Molecular dynamics          \\
\midrule
\multicolumn{4}{@{}l@{}}{\textit{Encapsulated (Enc)}} \\
\midrule
Athena++~\cite{athenapp}            & 132K  & C++           & Astrophysical MHD/AMR            \\
Smilei~\cite{smilei}                & 153K  & C++           & PIC plasma physics          \\
OpenLB~\cite{openlb}                & 338K  & C++           & Lattice Boltzmann CFD       \\
WarpX~\cite{warpx}                  & 553K  & C++           & PIC beam/plasma simulation          \\
\midrule
\multicolumn{4}{@{}l@{}}{\textit{Modularized (Mod)}} \\
\midrule
ROSS~\cite{ross}                    & 15K   & C             & Discrete-event simulation   \\
Nyx~\cite{nyx}                  & 441K  & C++           & AMR cosmology               \\
SAMRAI~\cite{samrai}            & 598K  & C++           & AMR mesh framework          \\
\bottomrule
\end{tabular}
\end{table}

\begin{figure}
  \centering
  \includegraphics[width=\linewidth, trim={142 150 155 140}, clip]{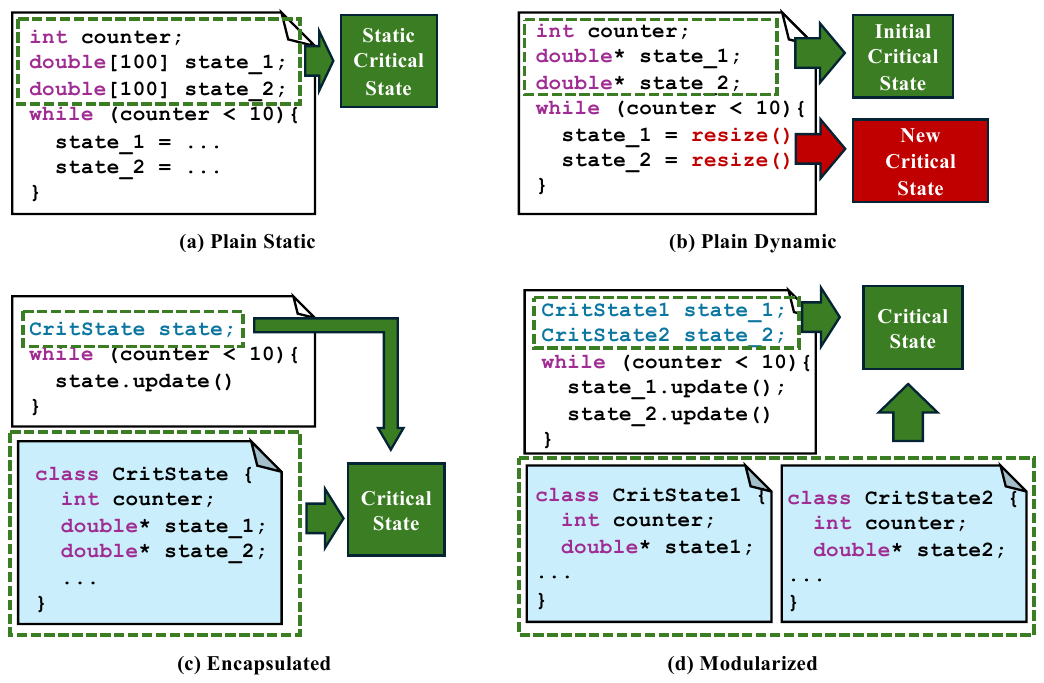}
  \caption{Applications classified by critical-state structure.}
  \label{fig:app-groups}
\end{figure}

\subsection{Critical State Classification}
To fill the gap above, we assemble a benchmark suite of 16 MPI scientific applications from diverse HPC domains. We select applications based on three criteria: (i) the source code is open and self-contained, so the agent is not limited by unavailable external context; (ii) applications include native checkpoint/restart implementations, which provide references for evaluating the generated code's correctness and efficiency; and (iii) application diversity such that our suite covers diverse domains, code sizes, computation and communication patterns, and critical-state structures, so that insights drawn from the suite are practically applicable. We group the applications into four classes, shown in \autoref{tab:apps} and \autoref{fig:app-groups}:

\begin{enumerate}[topsep=0pt,itemsep=0pt,leftmargin=*]
    \item \textbf{Plain Static (PS).} Critical state is stored in primitive data structures, such as scalars and arrays, that are directly managed by the application source code. The structure and shape of this state remain fixed throughout execution.
    
    \item \textbf{Plain Dynamic (PD).} Critical state is also stored in primitive, application-managed data structures, but its structure or ownership evolves over time, for example, when atoms migrate across MPI ranks.
    
    \item \textbf{Encapsulated (Enc).} Critical state is hidden behind a \textit{single} data structure or framework (e.g., AMReX~\cite{amrex}), exposed to the application only through predefined APIs.
    
    \item \textbf{Modularized (Mod).} Critical state is fragmented and distributed across \textit{multiple} application modules. Capturing and restoring this state requires understanding both the per-module state and the dependencies among modules.
\end{enumerate}
These groups represent increasing difficulty for checkpoint/restart. Plain static applications expose their critical state directly in the application code, making it relatively easy to identify what must be saved. Plain dynamic applications add the challenge of tracking how state changes over time and across MPI ranks. Encapsulated applications require the agent to reason through framework abstractions and, when necessary, write helpers to extract and restore framework-managed state. Modularized applications are the most challenging: the agent must understand multiple modules, their APIs, and dependencies to reconstruct a complete view of the application state and derive a correct checkpoint/restart strategy.

\subsection{Resilient Code-Refactoring Pipeline}


\begin{figure*}[hbtp]
    \centering
    \resizebox{\textwidth}{!}{%
\begin{tikzpicture}[
    font=\normalsize,
    >=Stealth,
    node distance=1.8cm and 1.8cm,
    doc/.style={
        draw=gray!90,
        fill=white,
        minimum width=1.7cm,
        minimum height=1cm,
        inner xsep=1pt,
        align=left,
        chamfered rectangle,
        chamfered rectangle corners={south east},
        chamfered rectangle xsep=2mm
    },
    agent/.style={
        draw=blue!60!black,
        fill=blue!70!black,
        text=white,
        minimum width=2.2cm,
        minimum height=1cm,
        align=center
    },
    validator/.style={
        draw=blue!60!black,
        fill=blue!70!black,
        text=white,
        minimum width=1.5cm,
        minimum height=1cm,
        align=center
    },
    llm/.style={
        draw=green!50!black,
        fill=green!70!black,
        text=white,
        minimum width=2.2cm,
        minimum height=1cm,
        align=center,
        font=\bfseries
    }
]

\node[doc] (ref) {\hspace*{-1mm}Reference\\\hspace*{-1mm}codebase};
\docfold{ref}
\node[doc, right=1.2cm of ref] (vanilla) {\hspace*{-1mm}Vanilla\\\hspace*{-1mm}codebase};
\docfold{vanilla}

\node[agent, right=1.1cm of vanilla] (agent) {Coding Agent};
\node[llm, above=0.6cm of agent] (llm) {LLM};

\node[doc, right=1.3cm of agent] (res) {\hspace*{-1mm}``Resilient''\\\hspace*{-1mm}codebase};
\docfold{res}

\node[validator, right=1.5cm of res] (validator) {Validator};
\node[validator, right=0.9cm of validator] (inspect) {Inspection};

\draw[->] (ref) -- node[above,align=center] {Remove\\[-1pt]resilient\\[-1pt]code} (vanilla);
\draw[->] (vanilla) -- node[above,align=center] {Initial\\[-1pt]prompt} (agent);
\draw[<->] (llm) -- (agent);
\draw[->] (agent) -- node[above,align=center] {Resilient\\[-1pt]code
\\[-1pt]injected} (res);
\draw[->] (res) -- node[above] {Validation} (validator);
\draw[->,green!60!black,thick]
    (validator) -- node[above,text=green!60!black] {PASS} (inspect);

\draw
    ($(validator.south)+(0.45,-0.35)$)
    node[text=red,font=\bfseries] {FAIL};
\coordinate (retrylevel) at ($(validator.south)+(0,-0.6)$);
\draw[->]
    (validator.south) |- (retrylevel)
    -- node[midway,above] {Retry}
       (agent.south |- retrylevel)
    |- (agent.south);
\end{tikzpicture}
}
    \caption{End-to-end agent-driven pipeline for adding checkpoint/restart resilience to an
             unmodified MPI application.}
    \label{fig:code-gen-process}
    \vspace{-15pt}
\end{figure*}
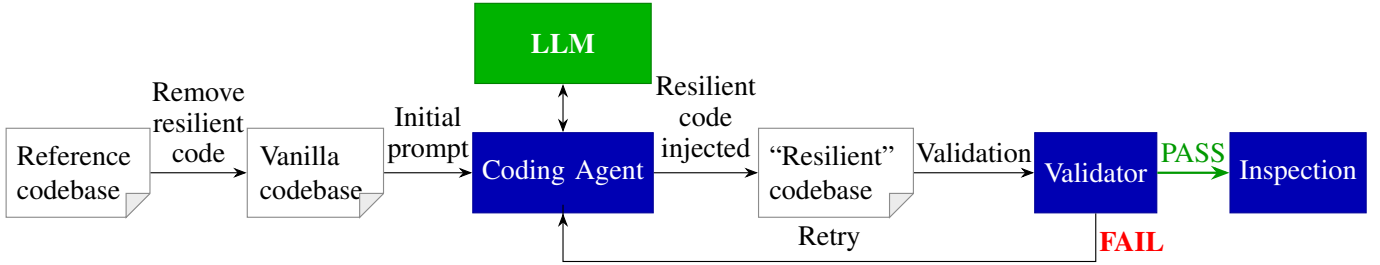

We design a no-human-in-the-loop code-refactoring pipeline, illustrated in \autoref{fig:code-gen-process}, to test whether frontier LLM coding agents can synthesize efficient checkpoint/restart implementations. Before giving an application to the agent, we remove its existing resilience logic to create a \emph{vanilla codebase}. \revise{This stripping preserves \textit{Reference}'s computation and failure-free scientific output but remains deliberately \textit{deep}}: it removes not only checkpoint/restart calls, but also framework-provided checkpoint read/write hooks. Thus, the agent cannot simply reuse or wrap the original implementation; it must infer the checkpoint/restart contract from the application's intrinsic data structures and execution logic.

The pipeline then runs an iterative generate--validate loop. In each iteration, the agent receives the original source code, \textit{Vanilla}, with a fixed prompt asking it to add checkpoint/restart support for fault tolerance. It uses tool calls to inspect and modify the source tree, producing a \emph{resilient codebase} that must preserve \textit{Vanilla}'s original functionality while adding checkpoint/restart capability. The prompt also requests the agent to log its reasoning, explain its tool calls, and plan for the next step after receiving tool call results and feedback. The log forms a per-iteration transcript that provides insight into the agent's thought for analysis. After the agent completes code generation, an independent validator then builds and tests the code.

\subsection{Validation Design}
In the pipeline, the validator is the key component for determining whether the generated code satisfies the resilience objective.
If so, the resilient codebase is promoted to the inspection phase, analyzed in \S\ref{sec:findings}. Otherwise, the agent receives structured feedback, including build logs, stdout/stderr, and per-condition pass/fail results to produce a revised implementation in the next iteration. 
The validator evaluates each resilient codebase along three dimensions: functionality, failure resilience, and trustworthiness.

\medskip
\noindent\textbf{Functionality.}
The resilient executable must preserve the original application behavior.
\revise{The validator compares the generated code’s failure-free output with \textit{Vanilla} using an application-specific tolerance $T$, calibrated from repeated runs to absorb floating-point nondeterminism while remaining far below differences across inputs.}

\medskip
\noindent\textbf{Failure resilience.}
The resilient code must mitigate failure impact by correctly recovering the application's recent state after failures. The validator tests this property by killing one MPI rank mid-run, relaunching the whole MPI run, and checking that: (i) the intended failure was actually delivered and at least one checkpoint was written before the kill, confirming that the resilient code truly uses checkpoint/restart for resiliency; (ii) the post-restart output matches that of \textit{Vanilla}, confirming that the checkpoint and recovery path reproduce the intended result; and (iii) the end-to-end wall time is meaningfully below a full restart-from-scratch execution, showing that the checkpoint cadence is operationally useful.

To make this test meaningful and tractable, each application input is tuned so the failure-free run lasts about 60--200 seconds, long enough for the application to perform substantial computation and produce meaningful output, while short enough to keep the validation time practical. Failures are injected uniformly between 50\% and 80\% of the failure-free runtime to prevent the agent from overfitting to a fixed kill point. This configuration is also a proxy for the failure-rate information available from the running environment to test whether the agent can use such information to choose an effective checkpoint frequency. The validator also enforces an upper bound on checkpoint size to fail whole-process memory dumps, forcing the agent to construct a checkpoint from its reasoning on the codebase.

\medskip
\noindent\textbf{Trustworthiness.}
The generated code must implement a genuine checkpoint/restart mechanism rather than gaming around the validator's feedback. This is important because prior studies showed that when frontier LLMs are rewarded for any change that makes the next validator run pass, they could produce code that merely exploits weaknesses in the validation process rather than solving the underlying problem~\cite{qi2015patch, zhong2026impossiblebench}. We observe similar behavior in our own loop. Across the 16 applications, the agent attempted at least once to find and copy the stripped framework checkpoint API inside the original codebase, cache the validator's expected output and replay it, ignore the checkpoint and re-execute the integrator post-failure, or checkpoint only a strict subset of the state that happened to dominate the comparator output.

To improve trustworthiness, the validator applies two defenses. 
First, it \textit{randomizes} inputs, making outputs unpredictable. This prevents the agent from replaying expected output to pass the validation instead of performing real recovery.
Second, it \emph{isolates} the code-refactoring and execution environment by cleaning data between tests and running the agent in a sandbox that limits its access to the resilient codebase and supporting data, such as log files. This prevents the agent from using tool calls to exploit results from previous runs, fabricate execution results, or copy solutions from the reference implementation.

\section{Benchmark Setup}
\label{sec:setup}

\medskip
\noindent\textbf{Objectives.}
We make the benchmark suite available at \url{https://github.com/ngduchai/autockpt}\footnote{Artifacts for reproducing the experiments in our key findings are available at \url{https://github.com/ngduchai/autockpt/blob/main/REPRODUCING.md}.} and use it to evaluate agent-driven checkpoint/restart generation along three dimensions. \textit{Feasibility:} can a coding agent add checkpoint/restart to support MPI applications without human intervention? \textit{Efficiency:} when it succeeds, how do the generated implementations compare with native human-written code? \textit{Challenges:} when it fails, what is the reason, and are the failures caused by application complexity, model limitations, or weaknesses in prompting, validation, and pipeline design?


\medskip
\noindent\textbf{Coding Agent Configuration.}
We choose \texttt{OpenCode} CLI~\cite{opencode} for the coding agent because it provides a rich tool interface for repository exploration and modification. We install VeloC~\cite{veloc}, a multi-level checkpoint/restart library for HPC, inside the \texttt{OpenCode} sandbox and provide its API and usage instructions in the input prompt. This relieves the agent from implementing low-level checkpoint I/O and lets it focus on resilience-specific tasks: identifying critical state, choosing safe checkpoint points, and implementing restart logic.
We evaluate three LLM backends:
\begin{enumerate}[topsep=0pt,itemsep=0pt,leftmargin=*]

    \item \textit{Claude Opus 4.7}. Anthropic's flagship model supports a 1M-token context window, priced at \$5/M input tokens and \$25/M output tokens~\cite{anthropicPricing}. Opus 4.7 is selected as a representative frontier LLM baseline.

    \item \textit{Claude Sonnet 4.6} provides a cheaper high-end LLM model with a 1M-token context window, priced at \$3/M input tokens and \$15/M output tokens~\cite{anthropicModels,anthropicPricing}. We use Sonnet 4.6 to evaluate whether a mid-tier yet cheaper model can generate resilient code, which could be beneficial for budget-constrained automation.

    \item \textit{GPT-5.5}, OpenAI's flagship model, provides a cross-provider comparison, with a 1.05M-token context window~\cite{openaiModels}, and pricing of \$5/M input tokens and \$30/M output tokens under standard-context usage; prompts above 272K input tokens use 2$\times$ input pricing and 1.5$\times$ output~\cite{openaiPricing}. We use GPT-5.5 for cross-provider comparison, representing the variability practitioners face when choosing among frontier coding models.
    
\end{enumerate}
For each application-model pair, the code-refactoring loop is capped at 10 iterations. Attempts that do not pass validation within this cap are reported as failures.

\medskip
\noindent\textbf{Metrics and Configurations.}
For each code generation attempt, the inspection phase records five metrics: (i) \emph{time cost}, including both code-refactoring and validator wall time; (ii) \emph{token cost}, input and output tokens across iterations; (iii) \emph{monetary cost} estimated from tokens used and official per-model API pricing; (iv) \emph{end-to-end execution time}, the wall time of one application run under each scenario, quantifying both instrumentation overhead and recovery cost; and (v) \emph{per-checkpoint footprint}, measuring the storage cost of a single checkpoint event.
All measurements are collected on a single development host using 4 MPI ranks per application, with three independent trials per configuration; we report the mean.

\section{Key Findings}
\label{sec:findings}

\begin{table}[t]
\centering
\caption{Iterations to generate a working checkpoint/restart implementation, reported as the median over successful attempts out of three trials. \fcolorbox{white}{green!20}{Numbers} give per-app lowest count across models;
\raisebox{0pt}{\fcolorbox{white}{red!20}{$\times$}} marks cases where all three attempts failed.
}
\label{tab:iterations}
\footnotesize
\setlength{\tabcolsep}{5pt}
\renewcommand{\arraystretch}{1.1}
\begin{tabular}{@{}llccc@{}}
\toprule
Group & App & Opus 4.7 & Sonnet 4.6 & GPT-5.5 \\
\midrule
\multirow{4}{*}{\textit{Plain Static}}
  & PRK\_Stencil & \cellcolor{green!20}2      & \cellcolor{green!20}2      & \cellcolor{green!20}2 \\
  & HPCG         & \cellcolor{green!20}1      & \cellcolor{green!20}1      & 4 \\
  & SW4lite      & \cellcolor{green!20}1      & 2                          & 5 \\
  & HyPar        & \cellcolor{green!20}2      & \cellcolor{green!20}2      & 3 \\
\midrule
\multirow{5}{*}{\textit{Plain Dynamic}}
  & CoMD         & \cellcolor{green!20}1      & \cellcolor{green!20}1      & 3 \\
  & SPPARKS      & \cellcolor{green!20}2      & \cellcolor{green!20}2      & \cellcolor{green!20}2 \\
  & CLAMR        & \cellcolor{green!20}3      & \cellcolor{green!20}3      & 6 \\
  & SPARTA       & 2                          & \cellcolor{green!20}1      & 4 \\
  & LAMMPS       & 5                          & \cellcolor{green!20}3      & \cellcolor{green!20}3 \\
\midrule
\multirow{4}{*}{\textit{Encapsulated}}
  & Athena++ & \cellcolor{green!20}3          & 5                          & \cellcolor{red!20}$\times$ \\
  & Smilei   & \cellcolor{green!20}1          & 2                          & 4 \\
  & OpenLB   & \cellcolor{green!20}2          & 3                          & 3 \\
  & WarpX    & 3                              & 4                          & \cellcolor{green!20}2 \\
\midrule
\multirow{3}{*}{\textit{Modularized}}
  & ROSS   & \cellcolor{green!20}5            & \cellcolor{red!20}$\times$ & \cellcolor{red!20}$\times$ \\
  & Nyx    & \cellcolor{red!20}$\times$       & \cellcolor{green!20}10     & \cellcolor{red!20}$\times$ \\
  & SAMRAI & \cellcolor{red!20}$\times$       & \cellcolor{red!20}$\times$ & \cellcolor{green!20}10 \\
\bottomrule
\end{tabular}
\end{table}

\begin{figure*}[t]
  \centering
  \begin{subfigure}[b]{0.24\textwidth}
    \centering
    \includegraphics[width=\linewidth]{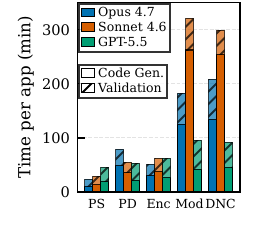}
    \caption{End-to-end wall time.}
    \label{fig:pergroup-time}
  \end{subfigure}\hfill
  \begin{subfigure}[b]{0.24\textwidth}
    \centering
    \includegraphics[width=\linewidth]{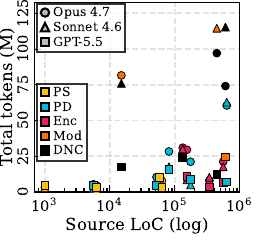}
    \caption{Tokens vs.\ source-tree size.}
    \label{fig:pergroup-loctokens}
  \end{subfigure}\hfill
  \begin{subfigure}[b]{0.24\textwidth}
    \centering
    \includegraphics[width=\linewidth]{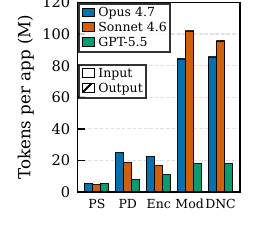}
    \caption{Token usage.}
    \label{fig:pergroup-tokens}
  \end{subfigure}\hfill
  \begin{subfigure}[b]{0.24\textwidth}
    \centering
    \includegraphics[width=\linewidth]{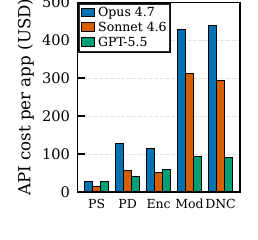}
    \caption{API monetary cost.}
    \label{fig:pergroup-cost}
  \end{subfigure}
  \caption{Code generation cost in terms of end-to-end wall time, token usage, and monetary cost characterized by application groups (PS = Plain Static, PD = Plain Dynamic, Enc = Encapsulated, Mod = Modularized, DNC = ``Did Not Complete'') and codebase size (Note: output tokens are negligible so they are almost invisible in \autoref{fig:pergroup-tokens})}
  \label{fig:pergroup}
  \vspace{-15pt}
\end{figure*}

\subsection{Feasibility of Enabling Resilience}
\label{sec:feasibility}

\autoref{tab:iterations} reports how many iterations each model needs to generate a working checkpoint/restart implementation. Across 16 applications and three LLM backends, the pipeline succeeds in 41 of 48 application--model settings. This result shows that agent-driven resilience generation is not limited to isolated cases: current coding agents can often synthesize working checkpoint/restart code across different domains, code sizes, and state structures without human intervention. To understand why these runs succeed, we inspect the per-iteration transcripts and find that agents consistently follow a five-stage workflow:

\begin{enumerate}[topsep=4pt,itemsep=2pt,leftmargin=12pt]

    \item \textbf{Reconnaissance.} The agent starts from the application entry point, such as \texttt{main.cpp}, and traces function calls inward until it locates the main iterative computation region, where execution state is updated and checkpoint/restart logic is most naturally inserted. It then scans build files such as \texttt{Makefile}, and creates a short to-do list that breaks the work into integration steps: prepare a configuration file for the runtime, update the build flags, identify critical state to save, place the save and restore calls, and validate.

    \item \textbf{Critical-state identification.} The agent uses the application's initialization path as the boundary between state that must be saved and state that can be regenerated. Variables initialized once from input files or reconstructable metadata are excluded, while variables updated inside the main iterative computation region are added to the checkpoint.

    \item \textbf{Checkpoint and restart placement.} The agent places checkpoint calls at a point where the application reaches a consistent state. Depending on the application iteration structure, this point could be after MPI synchronization (e.g., HPCG and CoMD), after diagnostics or output flushes (WarpX and OpenLB), or before the next communication phase (HyPar and LAMMPS). Restart logic is placed near the beginning of \texttt{main}, after runtime initialization but before compute kernels execute, guarded by a check for an available checkpoint. This structure closely matches the checkpoint/restart strategy found in many original human-written implementations, suggesting that the LLM's reasoning capability enables the coding agent to derive checkpoint/restart strategies that are structurally comparable to human-written implementations.

    \item \textbf{Source-tree integration.} The agent realizes its checkpoint/restart strategy using one of three implementation patterns. \emph{Inline}: injects the entire resilient code directly into the application code. \emph{Framework reuse}: reuses existing utilities (e.g., serialization) while implementing the solution to minimize code change impact. \emph{Helper-module}: creates a small module for checkpoint/restart logic and adds only a few calls to the module in the application code. GPT-5.5 tends to prefer helper modules, while Opus~4.7 and Sonnet~4.6 more often use inline integration and switch to framework reuse for encapsulated applications.
    
    \item \textbf{Validator-driven repair.} After the code integration, the agent passes the ``resilient-code'' to the validator and treats the validator's structured failure report (e.g., build error, output mismatch) as feedback. Each iteration reads the report, forms a single hypothesis about which part of the implementation is at fault, and proposes a targeted fix.
\end{enumerate}
This workflow is effective for most plain-static, plain-dynamic, and encapsulated applications. Opus~4.7 is the strongest overall: 11 of its 14 successful runs need the fewest iterations. Sonnet~4.6 requires fewer iterations than GPT-5.5, with a median of 3 iterations compared with 4. Across models, iteration count increases with checkpointing difficulty. The first iteration usually performs reconnaissance, state selection, placement, and integration; later iterations repair build errors, output mismatches, missing checkpoints from poor save cadence, or timeouts caused by faulty integration.

A limitation appears in modularized applications in which critical state is fragmented across modules, each exposing its own abstractions and hiding internal dependencies. To construct a correct checkpoint/restart implementation, the agent must infer which hidden fields are necessary for recovery, how those fields interact across modules, and how to restore them in a globally consistent way. This reasoning is difficult to perform without mistakes, and the validator feedback is often too coarse to guide systematic repair. In most cases, validation reports only the final symptom, such as an output mismatch or recovery failure, but not the underlying cause, such as which state was missing, which dependency was restored incorrectly, or which part of the restart path was broken.

As a result, the agent often enters a trial-and-error loop. It proposes a low-confidence hypothesis about the root cause, applies a local fix, and resubmits the code, but the next validation failure provides little additional information for narrowing the search, eventually trapping the agent until the iteration cap is reached. The few modularized successes, such as Opus~4.7 on ROSS, occur when the agent is tired of being trapped and abandons compact state selection, falling back to dumping all exposed module states. This fallback can produce a working implementation, but it is inefficient and unreliable: there is no guarantee that the agent will choose this strategy within the iteration budget, or that dumping exposed state is sufficient for every modularized application.

\vspace{2mm}
\noindent
\colorbox{blue!10}{
\parbox{0.96\linewidth}{
\underline{\textbf{Takeaway}:} \emph{
An LLM-driven coding agent can reliably implement checkpoint/restart solutions for applications whose critical state is visible or accessible through coherent abstractions, yet struggles when recovery state is fragmented.
}
}}

\begin{figure*}[t]
  \centering
  \includegraphics[width=0.95\textwidth]{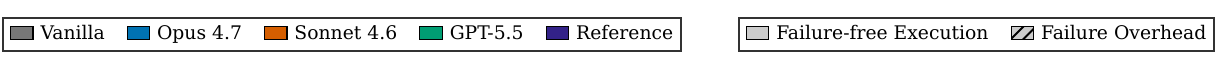}
  \vspace{-0.3em}

  \begin{subfigure}[b]{\textwidth}
    \centering
    \includegraphics[width=\textwidth]{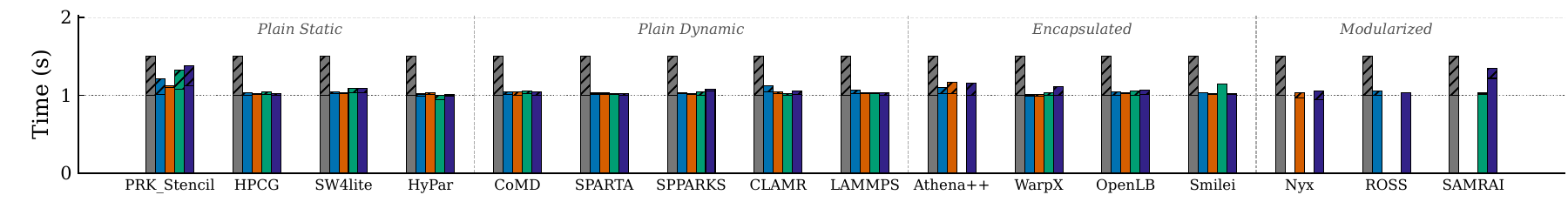}
    \setlength{\abovecaptionskip}{-15pt}
    \caption{Per-app execution time, normalized to \textit{Vanilla}'s failure-free runtime.}    
    \label{fig:walltime-overhead}
  \end{subfigure}

  \begin{subfigure}[b]{\textwidth}
    \centering
    \includegraphics[width=\textwidth]{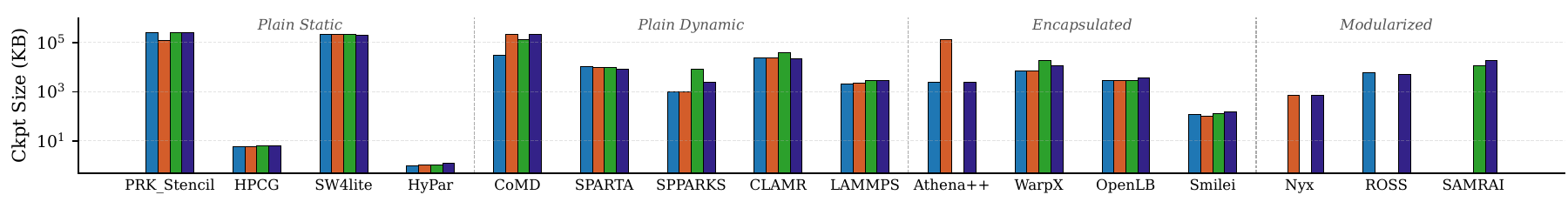}
    \setlength{\abovecaptionskip}{-15pt}
    \caption{Per-app complete checkpoint storage footprint.}
    \label{fig:ckpt-size}
  \end{subfigure}

  \caption{Efficiency of checkpoint/restart solutions generated by code agents in terms of (a) performance and (b) storage overhead. Reported values are medians across successful attempts. Missing bars indicate code-refactoring failures.}
  \label{fig:overhead}
  \vspace{-10pt}
\end{figure*}

\subsection{Code Generation Cost}
\label{sec:cost}

\medskip
\noindent\textbf{Time to Construct Resilient Code.}
We next evaluate the cost of producing resilient code. Across all attempts, including did-not-complete runs, the code-refactoring pipeline takes 73.4 minutes on average per application. The spread is large: the fastest application, HPCG, takes 19.7 minutes on average across models, while the slowest, Nyx, takes 217.5 minutes, a $\sim$11$\times$ difference. As shown in \autoref{fig:pergroup-time}, code generation and validation contribute comparable portions of wall time: code generation accounts for 58.5\% of total time on average, although the split varies by model. Sonnet~4.6 spends more time in code generation (70.6\%), whereas GPT-5.5 spends less (42.2\%), shifting more of its time to validation.

Generation time grows with checkpointing difficulty. The average attempt takes 32 minutes for plain-static applications, 61.7 minutes for plain-dynamic applications, 58.5 minutes for encapsulated applications, and 201.2 minutes for the three modularized successes, a 6.3$\times$ increase from the easiest to the hardest group. The increase is driven mainly by agent time: code generation alone rises from 14.1, 34.8, and 32.2 minutes for the first three groups to 141.9 minutes for modularized applications, comparable to the 131.7 minutes spent on did-not-complete attempts. Thus, the dominant cost is not simply editing code, but reasoning through increasingly difficult critical-state structures.

Model behavior differs substantially. Opus~4.7 is fastest overall, averaging 58.3 minutes compared with 61.5 for GPT-5.5 and 100.4 for Sonnet~4.6. Its advantage comes from fast completion: 11/14 of its successful runs pass validation with the fewest iterations, avoiding extra build, execution, and comparison cycles. GPT-5.5 requires more iterations overall, but each iteration is much cheaper in wall time: it spends only 4.5 minutes generating code per iteration, compared with 9.9 for Opus~4.7 and 12.8 for Sonnet~4.6. Consequently, in scenarios where all models require many iterations, such as modularized or did-not-complete cases, GPT-5.5 consumes roughly 2.5$\times$ less code-refactoring time per iteration.

\vspace{2mm}
\noindent
\colorbox{blue!10}{
\parbox{0.96\linewidth}{
\underline{\textbf{Takeaway}:} \emph{
Code-refactoring time scales primarily with critical-state complexity; strong models reduce time by completing in fewer iterations, while weaker models reduce the time spent in each iteration.
}
}}

\medskip
\noindent\textbf{Token and Monetary Cost.}
\autoref{fig:pergroup-loctokens} compares token usage with codebase size. Across applications ranging from 4K to 1M LoC, token usage remains mostly below 25M tokens, suggesting that cost does not grow directly with total codebase size. Agent transcripts explain why: during reconnaissance, the models first build a high-level picture of the source tree, then focus reading and editing on likely main iterative computation, allocation sites, and integration points instead of scanning the entire repository. This behavior is important for scalability: large codebases are not necessarily prohibitive when the agent can localize the relevant code.

Token usage instead follows critical-state complexity. Plain-static applications consume 1.79--9.97M tokens, averaging 5M per attempt. Relative to this baseline, plain-dynamic applications consume 3.4$\times$ more tokens, encapsulated applications 3.29$\times$ more, and modularized applications 12.4$\times$ more. The costliest attempt, Sonnet~4.6 on SAMRAI, reaches 115.3M tokens, producing a 64.5$\times$ range across the suite. Did-not-complete cases follow the same pattern: when the agent cannot infer the correct critical state, it remains trapped in repeated exploration and repair, burning tokens endlessly.
Input tokens dominate this cost. In \autoref{fig:pergroup-tokens}, input tokens account for 98.4\%--99.7\% of total usage. This reflects the nature of the task: the agent reads many source files and validation logs, but writes relatively small code patches. As a result, model behavior has a large effect on cost. Average token use per attempt is 22.8M for Opus~4.7, 30.4M for Sonnet~4.6 (133\% of Opus), and only 10.5M for GPT-5.5 (46\% of Opus, or about 2.2$\times$ fewer tokens).

This gap reveals an important trade-off. GPT-5.5 is much more token-efficient because it explores less code and commits to edits earlier. This makes it dramatically cheaper on simpler applications where the critical state is easy to identify. However, shorter exploration can also mean acting with less information. On applications with a more complex critical state, cheaper or more compact models may require more iterations, which increases both time and total cost. Sonnet~4.6 illustrates this tension: although it has lower per-token pricing than Opus~4.7, it can lose that advantage when additional iterations are needed, and on modularized applications, it runs slower and burns many more tokens than Opus.

\autoref{fig:pergroup-cost} converts token usage into monetary cost using the pricing in \S\ref{sec:setup}. Opus~4.7 costs \$116 per attempt on average, Sonnet~4.6 costs \$93 (20\% cheaper), and GPT-5.5 costs \$53.7 (54\% cheaper than Opus and 42\% cheaper than Sonnet). GPT-5.5 is therefore the clear cost leader; on modularized applications where it succeeds, it does so at roughly 1/4 the cost of the Anthropic models despite comparable iteration counts. However, the broader lesson is that model choice changes the cost profile substantially: deeper exploration can need less refinement, while compact exploration can reduce spending but risks extra repair iterations on difficult applications.

\vspace{2mm}
\noindent
\colorbox{blue!10}{
\parbox{0.96\linewidth}{
\underline{\textbf{Takeaway}:} \emph{
Generation cost is dominated by input tokens and scales with critical-state complexity rather than codebase size. Models that read less code can be dramatically cheaper, but this advantage diminishes when limited exploration causes extra iterations on complex applications.
}
}}

\subsection{Efficiency of the Generated Resilience Code}
\label{sec:efficiency}

\medskip
\noindent\textbf{Mitigating Failure Impact.}
\autoref{fig:walltime-overhead} reports execution time for LLM-generated resilient code, normalized to \textit{Vanilla}'s failure-free baseline. Each bar separates failure-free runtime from the extra time incurred after an injected failure. \textit{Vanilla} has no checkpoint logic and therefore loses all progress prior to the kill point.  The human-engineered \textit{Reference} uses the application's native checkpoint/restart implementation, while the LLM bars show code generated by Opus~4.7, Sonnet~4.6, and GPT-5.5. In failure-free execution, all LLM-generated implementations remain close to \textit{Vanilla}, usually within $3\%$, and are statistically indistinguishable from \textit{Reference}.

Under failure injection, the benefit is much larger. \textit{Vanilla} loses at least $50\%$ of its runtime because the kill is triggered after the execution reaches 50\% progress. In contrast, LLM-generated implementations usually confine failure impact to single-digit seconds, matching or improving on \textit{Reference}. Sonnet~4.6 produces the lowest failure impact on 8 applications, GPT-5.5 on 3 (HyPar, CLAMR, SPARTA), and Opus~4.7 on 2 (Athena++, WarpX).
\revise{Some generated implementations checkpoint more frequently and therefore lose less work after failure: on WarpX, they finish with about 1--3~s of impact versus 8.7~s for \textit{Reference}.}
These results show that the generated code is efficient: it exhibits recomputation overhead comparable with expert implementations.

\vspace{2mm}
\noindent
\colorbox{blue!10}{
\parbox{0.96\linewidth}{
\underline{\textbf{Takeaway}:} \emph{
Agent-driven code generation can mitigate the impact of failures at negligible overhead, comparable to human-implemented resilient solutions.
}
}}

\medskip
\noindent\textbf{Storage Overhead.}
\autoref{fig:ckpt-size} reports the size of one checkpoint snapshot for each LLM-generated implementation.
Plain-static and modularized applications are comparable to \textit{Reference}, but for different reasons. Plain-static state is simple, so the LLMs can select the same compact state as \textit{Reference}. Modularized applications are difficult enough that both \textit{Reference} and LLMs decide not to be selective, but conservatively dump the whole application state to the checkpoint, producing similar footprints. Plain-dynamic and encapsulated applications are more variable: the gap depends on whether the implementation checkpoints only restart-critical fields or saves a larger wrapper object. For example, on OpenLB \textit{Reference} dumps the full distribution-function lattice, while every LLM saves only the populations needed to resume the collision-streaming loop, producing smaller checkpoints. In contrast, on Athena++, Sonnet saves the full \texttt{MeshBlock} object while \textit{Reference} selects fields individually, producing checkpoints $\sim$53$\times$ larger (128 vs.\ 2.4~MiB).

Although the agents are not asked to optimize checkpoint size, they implicitly adopt this goal, starting with a minimal state and expanding it only if necessary. However, their selectivity is coarse-grained, i.e., entire variables are either included or excluded in the checkpoint. SPARTA illustrates this limitation: all three models save the full \texttt{Particle::OnePart} struct, including reconstructable or transient fields such as migration flags, writing 96~B per particle and roughly 9--10~MiB per snapshot. \textit{Reference} instead defines a restart-specific, 80~B per particle \texttt{Particle::OnePartRestart} struct, yielding a 7.8~MiB snapshot (about 1.2--1.3$\times$ smaller).

Model differences are also visible. Opus~4.7 stays within 1.3$\times$ of \textit{Reference} on every application where it inflates checkpoint size and produces the smallest checkpoint among the three LLMs on 10 of 16 applications. Sonnet~4.6 matches Opus on most applications but produces the Athena++ outlier (53$\times$) and falls behind on CoMD. GPT-5.5 inflates footprint relative to \textit{Reference} on most applications, making it the least space-efficient model.

\vspace{2mm}

\noindent
\colorbox{blue!10}{
\parbox{0.96\linewidth}{
\underline{\textbf{Takeaway}:} \emph{
LLM-generated implementations usually achieve storage overhead comparable to human-written code, but the footprint can vary widely.
}
}}

\section{Conclusion}
\label{sec:conclusion}

This paper studied whether frontier LLM coding agents can automate checkpoint/restart engineering for MPI scientific applications.
We show that agent-driven resilience engineering is feasible for a meaningful class of applications: across 16 applications, our no-human-in-the-loop pipeline produced 41 working implementations with negligible failure-free overhead and recovery efficiency comparable to human-written code. Success depends primarily on critical-state structure rather than codebase size. Agents are effective when recovery state is directly visible or accessible through coherent abstractions, but struggle when state is modularized, fragmented, and weakly diagnosed by validator feedback.
These findings suggest that agent-driven code generation can become a practical path toward lower-effort resilience, while also highlighting the need for future resilience-aware agents that combine code generation with state-dependency analysis, better source-code summarization, and more diagnostic validation feedback.

\revise{

\section*{Acknowledgment}
    
This work was supported in part by the U.S.\ Department of Energy under Contract DE-AC02-06CH11357 and National Science Foundation grants CSSI-2411386 and CSSI-2514056.

}

\balance

\bibliographystyle{IEEEtran}
\bibliography{references}

\end{document}